\documentstyle[multicol,aps,prl,epsf,psfig]{revtex}
\begin{document}
\title{Polydisperse Adsorption:~Pattern Formation Kinetics, Fractal 
Properties, and Transition to Order}
\author{N.~V.~Brilliantov$^{1,2}$, Yu.~A.~Andrienko$^{1,3}$,
P.~L.~Krapivsky$^{4}$, and J.~Kurths$^{3}$}
\address{$^{(1)}$Moscow State University, Physics Department,
Moscow 119899, Russia}
\address{$^{(2)}$Department of Chemistry, University
of Toronto, Toronto, Canada M5S 1A1}
\address{$^{(3)}$University of Potsdam, 
Physics Department, Am Neuen Palais, 
D--14415 Potsdam, Germany}
\address{$^{(4)}$ Center for Polymer Studies and Department of Physics,
Boston University, Boston, MA 02215}
\maketitle

\begin{abstract}
  
  We investigate the process of random sequential adsorption of polydisperse
  particles whose size distribution exhibits a power-law dependence in the
  small size limit, $P(R)\sim R^{\alpha-1}$.  We reveal a relation between
  pattern formation kinetics and structural properties of arising patterns.
  We propose a mean-field theory which provides a fair description for
  sufficiently small $\alpha$.  When $\alpha \to \infty$, highly ordered
  structures locally identical to the Apollonian packing are formed.  We
  introduce a quantitative criterion of the regularity of the pattern
  formation process.  When $\alpha \gg 1$, a sharp transition from irregular
  to regular pattern formation regime is found to occur near the jamming
  coverage of standard random sequential adsorption with monodisperse size
  distribution.

\end{abstract}

\begin{multicols}{2}

\section{Introduction}

Random sequential adsorption (RSA) is an irreversible process in which
particles are adsorbed sequentially and without overlaps and deposited
particles cannot diffuse or desorb from the substrate.  The RSA model
has been initially applied to reactions along polymer
chains\cite{flory}.  More recently, RSA processes have found a variety
of other applications from adhesion of colloidal particles and proteins
onto substrates\cite{fed} to chemisorption\cite{chem} and epitaxial
growth\cite{weeks}; for reviews see\cite{rev}.

In this paper, we study a simple generalization of RSA that creates a rich
dynamic behavior and results in complex spatial patterns.  Namely, we
consider polydisperse random sequential adsorption (PRSA) processes.
Adsorption of mixtures has been addressed in a very few
studies\cite{baker,talbot,meakin,pk}.  If a mixture contains a small number
of different sizes, geometric and kinetic characteristics are primarily
determined by the smallest size.  In many applications, however, the size
distribution is continuous and spreads over several decades\cite{powder}.
Therefore, before the smallest size will finally win, an interesting
intermediate asymptotics arises.  To address this intermediate regime, we
consider PRSA with the power-law distribution in the small size limit,
$P(R)\sim R^{\alpha-1}$.  We show\cite{bakk} that this PRSA gives rise to
fractal patterns of dimension $D_f$ depending on the exponent $\alpha$.
Additionally, we measure the degree of order of the patterns formed by PRSA,
and identify the local structure of the patterns arising in the limit
$\alpha\to\infty$ with Apollonian packing\cite{mand}.  The significance of
Apollonian packing in surface deposition problems was also recognized in
Ref.\cite{other}. 

This paper is organized as follows. In Sec.~II, we introduce PRSA and present
numerical results.  In Sec.~III, we develop scaling, exact, and mean-field
approaches to PRSA.  Exact results are available for the one-dimensional (1D)
PRSA, and they are used to check mean-field and scaling approaches.  A
relation between kinetics and geometry of arising patterns is also discussed.
In Sec.~IV, we introduce a quantitative criterion of regularity of the
pattern formation process and analyze the ordering in PRSA processes.  The
last Sec.~V contains a summary.

\section{Polydisperse RSA}

In applying RSA to real processes, one should take into account that
adsorbed particles are typically polydisperse.  The relevant example
is adsorption of colloidal particles, that have a broad radii
distribution.  It is usually described by the Schulz
distribution\cite{shulz}, which has a power-law dependence on the
radius $R$ for small $R$ and an exponential tail for large $R$:
\begin{equation}
P_{Sz}(R)=\left[ \frac{\alpha}{\langle R\rangle}\right]^{\alpha} 
\frac {{}R^{\alpha-1}}{\Gamma(\alpha)}
\exp \left[-\frac{\alpha}{\langle R\rangle}R\right].
\label{disshulz}
\end{equation}
Here $\langle R\rangle$ is the average radius
and $\Gamma(x)$ the gamma function.  Note that the exponent $\alpha$
should be positive to obey the normalization requirement,
$\int dR\,P(R)=1$.

Only the small-size behavior of $P(R)$ affects the most interesting long time
characteristics since in this regime only small particles can be adsorbed.
Thus instead of (\ref{disshulz}) we shall use a power-law size distribution
with an upper cutoff (taken as the unit of length):
\begin{eqnarray}
\label{distrib}
P(R)=\cases{\alpha \, R^{\alpha -1} & $R\leq 1$;\cr
                                  0 & $R>1$.}
\end{eqnarray} 

The patterns formed by PRSA are drastically different from traditional RSA
patterns, since the coverage is complete for PRSA.  The pore space of the
patterns is a nontrivial fractal set.  This is physically evident, and in one
dimension it proves possible to determine the fractal dimension $D_f(\alpha)$
analytically .  In higher dimensions, we have to resort to numerical
treatment.

Monte Carlo simulations of the PRSA model have been performed by implementing
the following algorithm.  A center of the new disc is chosen at random with a
uniform probability density.  The radius of the disc is generated according
to the size-distribution of Eq.~(\ref{distrib}).  If this disc does not
overlap any other discs already in placed, it is deposited.  Otherwise, the
attempt is discarded.  The maximal coverage studied in simulations was $\sim
0.9$ and about $100$ runs were performed for each $\alpha$.  Some of the
generated patterns are shown in Fig.~1.  Clearly, the character of patterns
changes when the exponent $\alpha$ increases from $0$ to $\infty$.  For the
small $\alpha$, the patterns look like a random set of little discs
distributed uniformly over the plane, with larger discs randomly scattered in
the ``sea'' of smaller ones.  For large $\alpha$, one recognizes a structure
initially formed by large discs and then reproduced by smaller discs in the
holes between the large ones.  These properties of patterns follow from the
small size behavior of radii distribution function $P(R)$.  Indeed when
$\alpha\to 0$ the smallest particles primarily participate in the adsorption,
and hence apparently random patterns emerge.  When $\alpha \to\infty$, the
particles of the largest size are deposited until the system reaches the
jamming limit of monodisperse RSA.  Then the next particle to arrive will be
the one that fits the biggest hole.  This continues, so in this second stage
the process is deterministic and apparently ordered patterns emerge.

On the length scales much smaller than the upper cutoff, patterns are
self-similar (Fig.~2). This suggests the fractal nature of the arising
patterns.  We have measured the fractal dimension $D_f$ of the
pore-space of the patterns using the standard approach (see e.g.
\cite{manna,ab}) based on the analysis of the radii distribution
function for the adsorbed particles.  Namely, denote by $n(R)dR$ the
number density of adsorbed discs with radii from the interval
$(R,R+dR)$.  Let $\epsilon$ is an arbitrary lower cutoff radius.  Then
$N(\epsilon) = \int_{\epsilon} ^{\infty} dR\,n(R)$ gives the number
density of discs with radii greater than $\epsilon$.  The power law
behavior of $N(\epsilon)$ at the $\epsilon \to 0$ limit,
\begin{equation}
N(\epsilon) \sim \epsilon ^{-D_f},
\label{frdim}
\end{equation}
is a signature that the pore space is a set of fractal dimension $D_f$.
Numerically, we indeed observed this power law behavior.  We also found
that $D_f$ monotonously decreases when $\alpha$ increases (see Fig.~3).

\section{Theoretical approaches to pattern formation}

In the present section we employ scaling, mean-field and exact
approaches to the process of pattern formation in PRSA. The first
approach is based on the scaling hypothesis and gives relations
between structural and kinetic characteristics of the system. 

\subsection{Scaling Framework}

Let $\Phi(t)$ is the fraction of uncovered area and $\Psi(R,t)$ the
probability that a disc of radius $R$ can be placed onto a
substrate\cite{bakk}.  Clearly,

\begin{equation}
\Psi(0,t)=\Phi(t).
\label{psiphi}
\end{equation}
$\Phi(t)$ evolves according to the {\em exact}
rate equation

\begin{equation}
\frac{d\Phi }{dt}=-\int_0^\infty dR P(R)\Psi (R,t)V_d R^d,
\label{phi1}
\end{equation}
where $V_d$ denotes the volume of the $d$-dimensional unit
sphere, $V_d=\pi^{d/2}/\Gamma(1+d/2)$.  
Assuming a  scaling behavior of $\Psi(R,t)$,  we write

\begin{equation}
\Psi (R,t)=S^{\theta}(t)\,F\left(\frac R{S(t)}\right).
\label{psi2}
\end{equation}
Here $S(t)\sim t^{-\nu }$ is a typical gap between neighboring adsorbed
particles and $F(x)$ is a scaling function.  The scaling description
applies when $t\to\infty$ and $R\to 0$ with $R/S(t)$ finite.  The
existence of scaling is an assumption, which is supported by numerical
evidence in 2D and by analytical results in 1D\cite{pk}.

Eqs.~(\ref{psiphi}) and (\ref{psi2}) imply $\Phi (t)\sim t^{-z}$ with
$z=\theta\nu$. Substituting then (\ref{psi2}) into (\ref{phi1}) gives
\begin{equation}
\nu=\frac{1}{\alpha+d}, \quad
z \simeq \alpha V_d\int_0^\infty dx\, x^{\alpha+d-1} F(x).
\label{nuz}
\end{equation}

Scaling suggests that self-similar fractal structures should arise.
Indeed, computing the number density of the absorbed particles we get
\begin{equation}
n(R)=\int_0^\infty dt\,P(R)\Psi (R,t) \sim R^{\alpha-1+(z-1)/\nu }
\label{nr}
\end{equation}
Hence, the power-law dependence of Eq.~(\ref{frdim}) is recovered
when we identify the fractal dimension with
\begin{equation}
D_f=d-z(d+\alpha).
\label{df}
\end{equation}
Eq.~(\ref{df}) provides a relation between the fractal dimension of the
arising patterns and the kinetic exponent $z$.  Similar qualitative
behaviors were observed in other pattern formation models\cite{ab}.

We should stress that scaling provides just a framework; for instance, it
gives scaling relations among the exponents but it does not allow to compute
the exponents.  So one should use other approaches to get a complete exact or
approximate description of PRSA.  In one dimension, an exact description is
indeed possible. In higher dimensions, even in the extreme case of
$\alpha=\infty$ the fractal dimension of the pore space remains unknown, so
analytical description of PRSA is hardly possible.

\end{multicols}

\begin{minipage}{16.0cm}
\begin{figure}[htbp]
  \centerline{
  \psfig{file=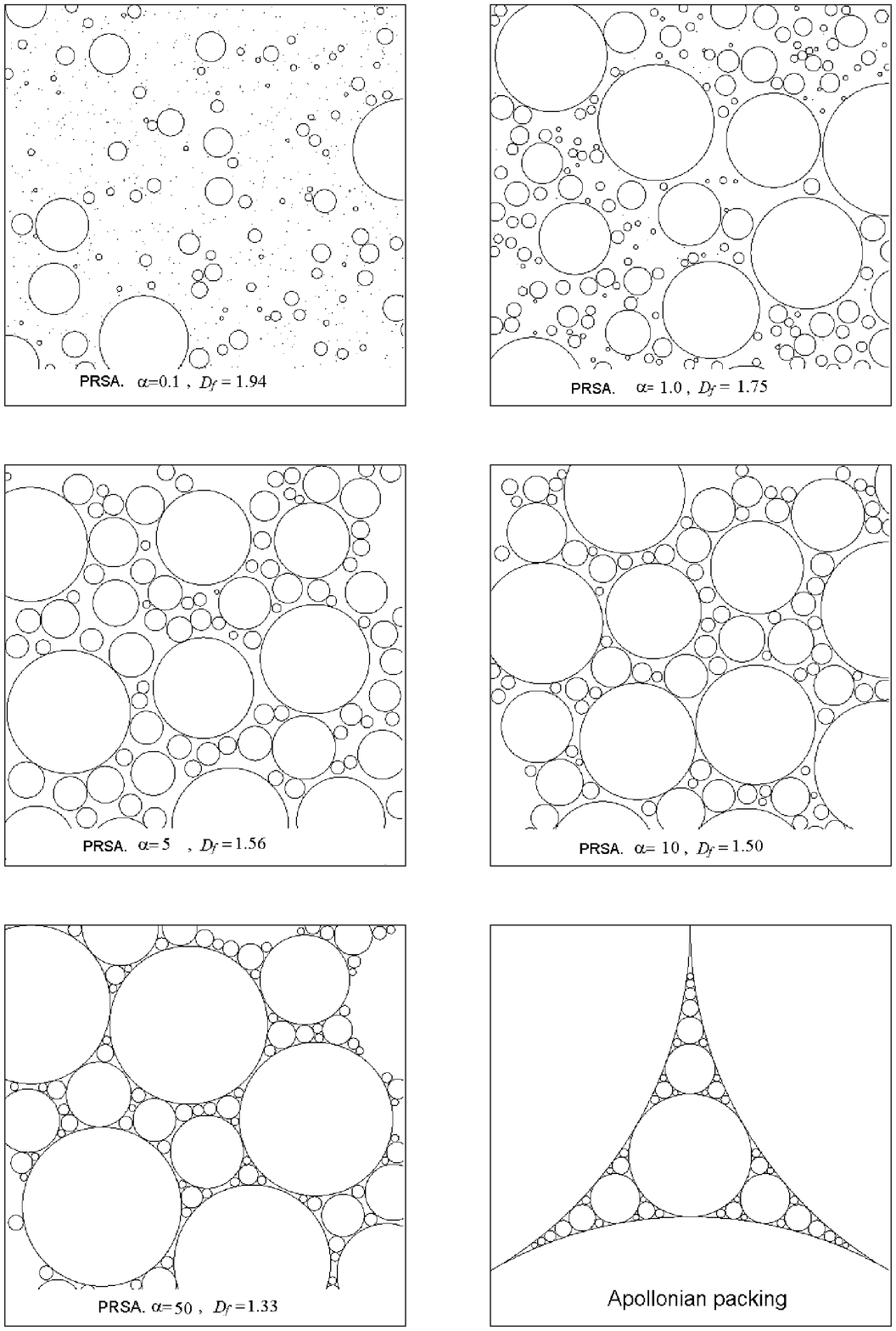,width=15.8cm}}
\vspace{0.4cm}
\noindent{{\bf Fig.~1} Typical PRSA patterns for $\alpha=0.1, 1, 5, 10, 50$ and
Apollonian packing.}
\label{fig:1}
\end{figure}
\end{minipage}

\begin{minipage}{16.0cm}
\begin{figure}[htbp]
  \centerline{
  \psfig{file=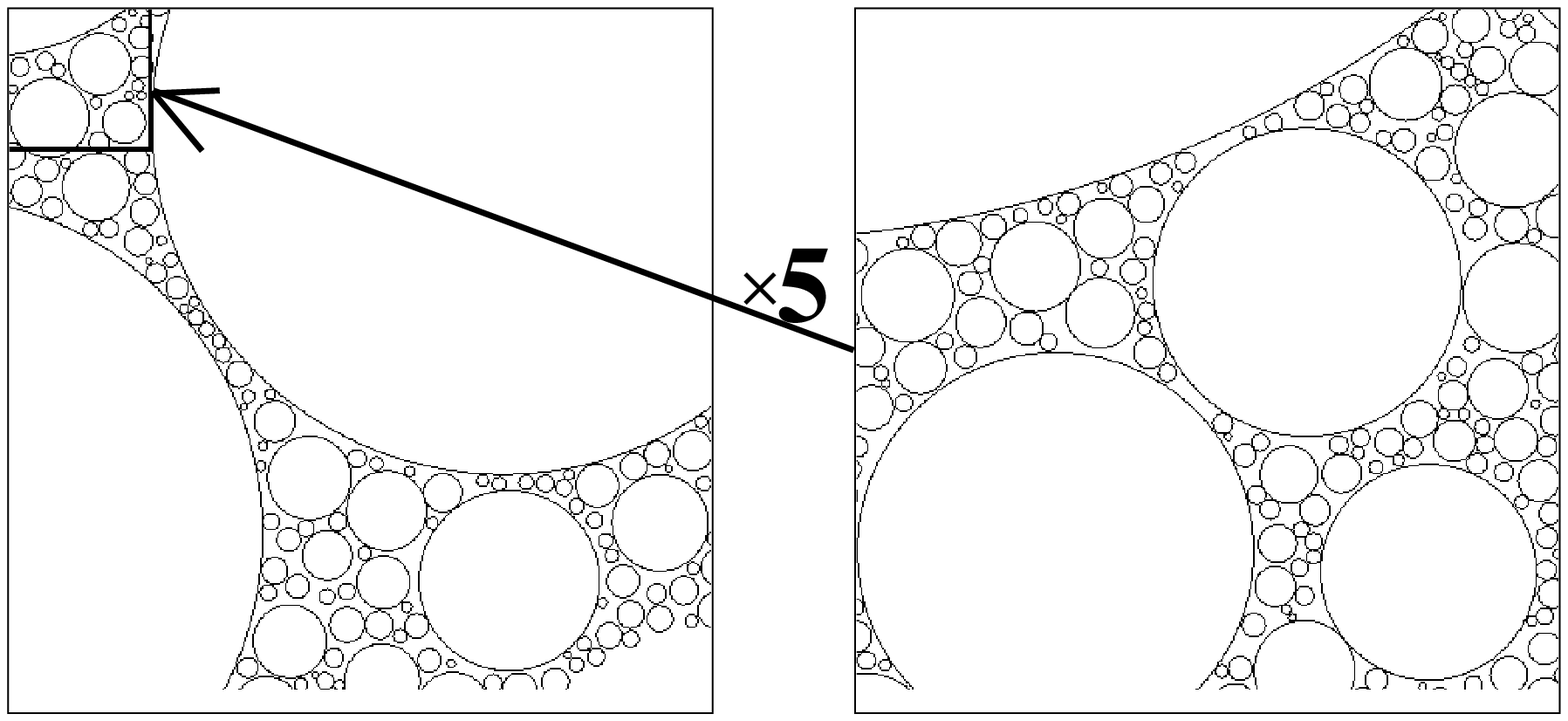,width=15.8cm}}
\vspace{0.4cm}
\noindent{{\bf Fig.~2} Self-similarity of the PRSA patterns.}
\label{fig:2}
\end{figure}
\end{minipage}
\vspace{0.1cm}

\begin{multicols}{2}

\subsection{1D PRSA: Exact Results}

A detailed analysis of 1D PRSA is given in Ref.~\cite{pk}.  Here we sketch
basic results which are necessary to determine the fractal dimension.  Let
$C(x,t)$ be the concentration of holes of length $x$ at time $t$.  All holes
have the same "shape" in 1D that significantly simplifies the problem.  By
definition, $C(x,t)$ is related to $\Psi(x,t)$ via

\begin{equation}   
\Psi (x,t)=\int_x^{\infty} dy (y-x)C(y,t).
\label{chipsi}
\end{equation}
The kinetic equation for $C(x,t)$ reads

\begin{eqnarray}
\frac{\partial C(x,t)}{\partial t}=
-C(x,t) \left[ \int_0^x dz(x-z)P(z)\right] \nonumber\\
+2 \int_x^{\infty}dy\,C(y,t) \int_0^{y-x} dz\,P(z). 
\label{1dgen}
\end{eqnarray}
The first term on the right-hand side of Eq.~(\ref{1dgen}) gives the loss of
holes of length $x$ due to deposition of intervals of length $z$ (with
$z<x$); the second term describes the gain of holes of length $x$ from larger
holes.  Substituting $P(z) \equiv P(R)$ given by Eq.~(\ref{distrib}) into
Eq.~(\ref{1dgen}) yields

\begin{equation}
\left(\frac{\partial}{\partial t}+\frac{x^{\alpha +1}}{\alpha+1}\right)
C(x,t)=2\int_x^{\infty}dy\,C(y,t) (y-x)^{\alpha}.
\label{1dkin}
\end{equation}
Multiplying both sides of Eq.~(\ref{1dkin}) by $x^{\beta}$ and
integrating over $x$ gives the kinetic equation
\begin{equation}   
{dM_{\beta}\over dt}=\left[ 
2 \frac{\Gamma(\alpha +1)\Gamma(\beta +1)}{\Gamma(\alpha +\beta +2)}-
\frac{1}{\alpha +1} \right] M_{\alpha + \beta +1}
\label{momeq}
\end{equation}
for the moments of the hole-size distribution
\begin{equation}
M_{\beta}(t)=\int_0^{\infty} dx\,x^{\beta} C(x,t).
\end{equation}
Eq.~(\ref{chipsi}) implies 
$C(x,t)={\partial^2\over \partial x^2}\,\Psi(x,t)$.  Combining this with
the scaling ansatz of Eq.~(\ref{psi2}) we obtain
\begin{equation}
\label{csf}
C(x,t)=S^{\theta-2}(t)\,F''\left(x/S(t)\right),
\end{equation}
where $F''(\xi)=d^2F/d\xi^2$. 
Eq.~(\ref{csf}) allows us to express the moments $M_{\beta}(t)$ 
via the time-independent moments,
\begin{equation}
M_{\beta}(t)=S^{\theta+\beta-1}(t) m_\beta, \quad
m_{\beta}=\int_0^{\infty} d \xi\,\xi^{\beta} F''(\xi).
\label{moms}
\end{equation} 
Choose now the exponent $\beta$ so that the numerical factor on the
right-hand side of Eq.~(\ref{momeq}) vanishes, i.e.,
\begin{equation}
2 \frac{\Gamma(\alpha +1)\Gamma(\beta +1)}{\Gamma(\alpha +\beta +2)}=
\frac{1}{\alpha+1}.
\label{eqbeta}
\end{equation}
For such $\beta=\beta(\alpha)$, Eq.~(\ref{momeq}) implies that $M_{\beta}(t)$
does not depend on $t$.  Eq.~(\ref{moms}) therefore gives $\theta=1-\beta$,
and then other exponents and the fractal dimension are found:
\begin{equation} 
\nu=\frac{1}{1+\alpha}, \qquad
z=\frac{1-\beta}{1+\alpha}, \qquad
D_f=\beta.
\label{zbeta}
\end{equation}
A simple analysis shows that (i) Eq.~(\ref{eqbeta}) has only one positive
solution $\beta=\beta(\alpha)$, (ii)~$\beta<1$ for all $\alpha>0$, and
(iii)~$\beta$ decreases when $\alpha$ increases.  One can determine
$\beta=D_f$ explicitly in some specific cases, e.g., $D_f=(\sqrt{17}-3)/2$
for the uniform size distribution ($\alpha=1$).

\subsection{Mean-Field Theory of PRSA}

In higher dimensions we employ an approximate mean-field treatment.  We shall
use a mean-field description of PRSA close to the one developed in\cite{ab}
for nucleation-and-growth processes.  As usual, we shall ignore
many-particles spatial correlations and account only two-particle ones.  To
find $\Psi(R,t)$, consider a circle of radius $R$ centered at the origin.
Then one can write the following estimate to this function in the mean-field
spirit:
\begin{eqnarray}
\label{psimft}
&&\Psi(R,t)= \nonumber\\
&&\exp \left\{-\int_0^t d\tau \int_0^R dr\,\Omega_d 
r^{d-1}\int_0^\infty d\rho\,P(\rho){\Psi(\rho,\tau)\over
\Phi(\tau)}\right\}\\ 
&&\times \exp \left\{-\int_0^td\tau \int_R^\infty dr\,\Omega_d r^{d-1}
\int_{r-R}^\infty d\rho\,P(\rho )\frac{\Psi(\rho,\tau )}
{\Phi(\tau)}\right\}. \nonumber  
\end{eqnarray}
The former exponential factor in (\ref{psimft}) estimates the
probability that our circle is not covered up to time $t$ by discs with
centers fall inside this circle. The latter exponential factor
guarantees that the free space inside our circle is not covered by the
other discs which were adsorbed outside the circle.  We also denote by
$\Omega_d$ the surface area of unit sphere in $d$ dimensions,
$\Omega_d=dV_d=2\pi^{d/2}/\Gamma(d/2)$.  The exponential factors in
Eq.~(\ref{psimft}) were derived by noting that
\begin{displaymath} 
d\tau\,dr\,d\rho\,\Omega _d r^{d-1}P(\rho)\Psi(\rho,\tau)  
\end{displaymath}
gives the probability that a disc of radius belonging to
the interval $(\rho,\rho+d\rho)$ was adsorbed 
within the time interval $(\tau ,\tau+d\tau)$ in the spherical shell
centered at the origin and confined by radii $r$ and $r+dr$.  The
factor $\Psi(\rho,\tau)$ guarantees that the disc of radius $\rho$
may be placed into the system.  Such event prevents the adsorption of
disc of radius $R$ at the origin. The probability that such event has
not happened,
\begin{displaymath} 
1-d\tau\,dr\,d\rho\,\Omega_dr^{d-1} P(\rho )\Psi (\rho,\tau),
\end{displaymath}
may be re-written as
\begin{displaymath} 
\exp\left\{-d\tau\,dr\,d\rho\,\Omega_dr^{d-1} P(\rho )\Psi (\rho,\tau)
\right\}.  
\end{displaymath}
The probability that none of these events has happened up to time $t$ is
obtained by multiplying all these factors with $0\leq \tau \leq t$,
$0\leq r\leq \infty$, and $0\leq \rho \leq \infty$.  However one should
take into account that above expression gives a correct estimate only
for factors with $\tau=0$. For subsequent factors with $\tau>0$, one
should use $\Psi(\rho ,\tau )/\Phi(\tau)$ instead of $\Psi(\rho,\tau)$
since the preceding factors guarantee that the discs are placed on the
uncovered space.  Using $\Phi(0)=1$ and treating separately $r\leq R$
and $r\geq R$, one arrives at Eq.~(\ref{psimft}).  Again we assume that
in the scaling regime we can use Eq.~(\ref{psi2}) for the function
$\Psi(R,t)$.  Taking into account that $\Psi(0,t)=\Phi(t)$, we get
\begin{eqnarray}
\label{eqf}
&&F(Rt^{\nu})= \\
&&\exp \left\{ -\int_0^td\tau \int_0^\infty d\rho\,
\Omega_dP(\rho )\left((R+\rho )^d
-\rho ^d\right) F(\rho\tau^{\nu})\right\}.  \nonumber
\end{eqnarray}
Remarkably, the ansatz
\begin{equation}
F(x)=\exp(-A_1x-\ldots -A_dx^d)
\label{fanz}
\label{fx}
\end{equation}
reduces the nonlinear integral equation (\ref{eqf}) to a system of $d$
algebraic equations for coefficients $A_j$. In particular, in 1D we get
$\nu=1/(\alpha+1)$ and a closed equation for $A_1$:
\begin{equation}
\label{A1}
A_1=\frac{2\alpha}{1-\nu \alpha}
\int_0^{\infty}dx\,x^{\alpha-1}e^{-A_1x}.
\end{equation} 
Eq.~(\ref{A1}) is solved to find $A_1=\left[2\Gamma(\alpha+2)
\right]^{\frac{1}{\alpha+1}}$.  Thus on the mean-field level, the
scaling function for 1D PRSA is pure exponential, i.e.  it is clearly
different from the analytical solution.  To obtain a quantitative
difference we substitute $F=e^{-A_1x}$ with the above value of $A_1$
into Eq.~(\ref{nuz}) to find $z=\frac{\alpha}{\alpha+1}$.  Substituting
this into Eq.~(\ref{df}) we obtain the fractal dimension: $D_f^{\rm
MF}=1-\alpha$ for $\alpha<1$ and $D_f^{\rm MF}=0$ for $\alpha\geq 1$.
Thus the mean-field theory is clearly wrong for $\alpha\geq 1$, though
in the limit $\alpha\to 0$ it becomes exact\cite{rem102}.  For instance,
$D_f^{\rm exact}-D_f^{\rm MF} =(\frac{2\pi}{3}+4\gamma^2-2-6\gamma)
\alpha^2 + {\cal O}(\alpha^3)$ (here $\gamma\cong 0.57721566$ is
the Euler constant).

For 2D PRSA, the ansatz of Eq.~(\ref{fanz}) yields 
\begin{eqnarray*}
A_{1}&=&2 \pi\alpha(\alpha +2) \int_0^{\infty}dx\, 
x^{\alpha} e^{-A_1x-A_2x^2}, \\
A_{2}&=&\frac{\pi}{2} \alpha(\alpha +2)\int_0^{\infty} dx\,
x^{\alpha-1} e^{-A_1x-A_2x^2}.
\end{eqnarray*}
Solving these equations numerically, and then inserting
$F(x)=e^{-A_1x-A_2x^2}$ into Eq.~(\ref{nuz}), one finds $z$ and $D_f$.

In the small $\alpha$ limit, we perform a perturbation analysis to 
find\cite{rem5}
\begin{eqnarray*}
A_{1}&=&2\pi\alpha+a_1\alpha^2
=2\pi\alpha-17.3557\alpha^2,\\
A_{2}&=&\pi+a_2\alpha + a_3\alpha^2=
\pi-1.8339\alpha +12.6546\alpha^2,
\end{eqnarray*} 
where we have omitted terms of order ${\cal O}(\alpha^3)$.  Using these
expressions for $A_{1}$ and $A_{2}$, we get the kinetic exponent
\begin{equation}
\label{zmf}
z=\alpha/2+a_4\alpha^2=\alpha/2-1.5428\alpha^2+\cdots
\end{equation}
and for the fractal dimension
\begin{equation}
\label{dmft}
D_f=2-\alpha+a_5\alpha^2=2-\alpha+2.5856\alpha^2+\cdots. 
\end{equation}

The mean-field predictions for $D_f(\alpha)$ and $z(\alpha)$ are shown on
Fig.~3 and Fig.~4.  We see that the mean-field approach provides a fair
description for small $\alpha$.  For $\alpha \simeq 1$, however, the spatial
correlations become more and more important and the mean-field theory fails.

\begin{figure}
%\narrowtext
\epsfxsize=2.7in\epsfysize=2.0in
\hskip 0.3in\epsfbox{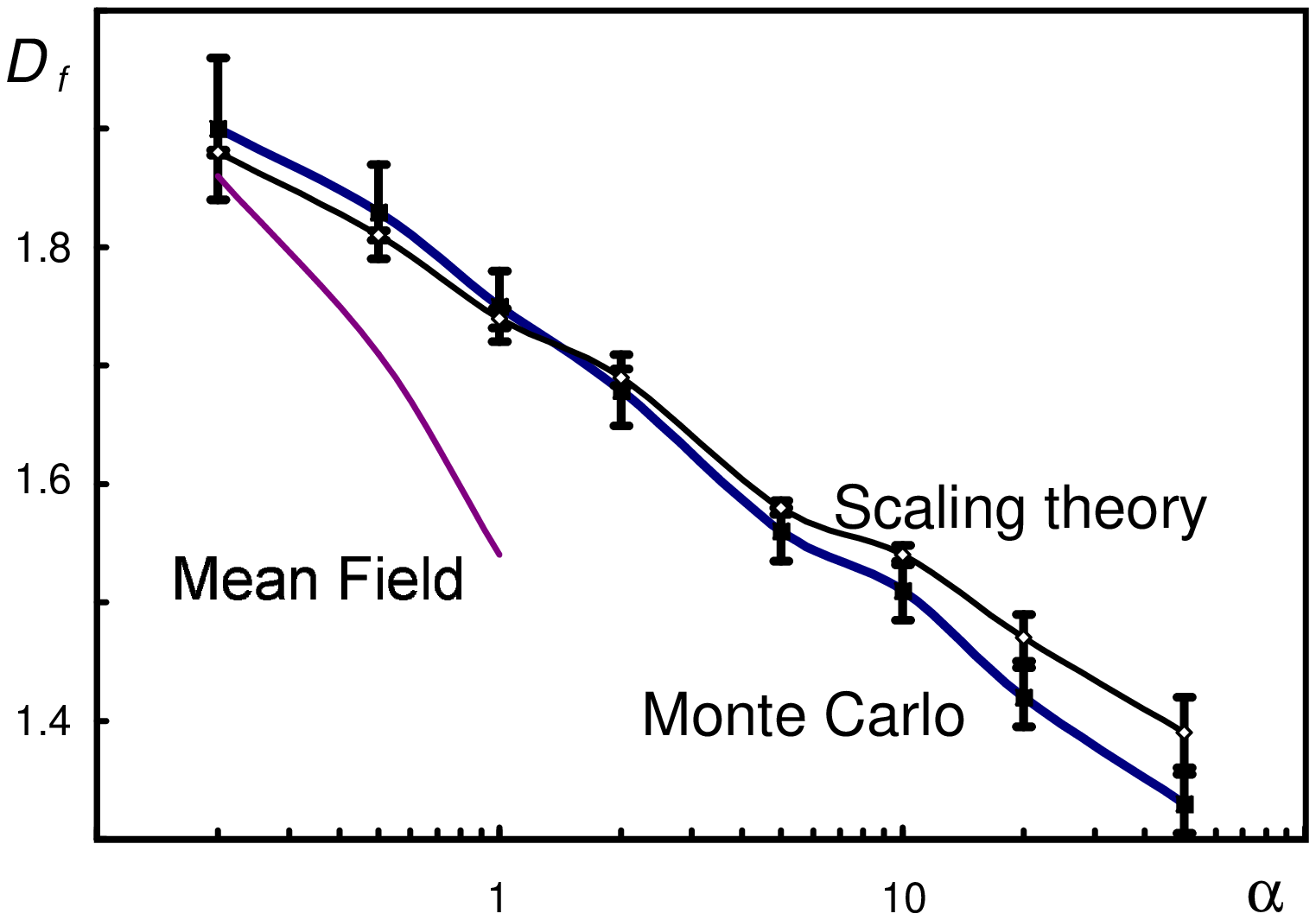}
\vskip 0.15in
\noindent{{\bf Fig.3}
The fractal dimension $D_f$ vs. $\alpha$ in two dimensions. 
$D_f$ of the scaling theory is obtained from Eq.~(\ref{df}) with the 
``experimental'' value of the kinetic exponent $z$.}
\label{fig:3}
\end{figure}

\begin{figure}
%\narrowtext
\epsfxsize=2.8in\epsfysize=2.35in
\hskip 0.3in\epsfbox{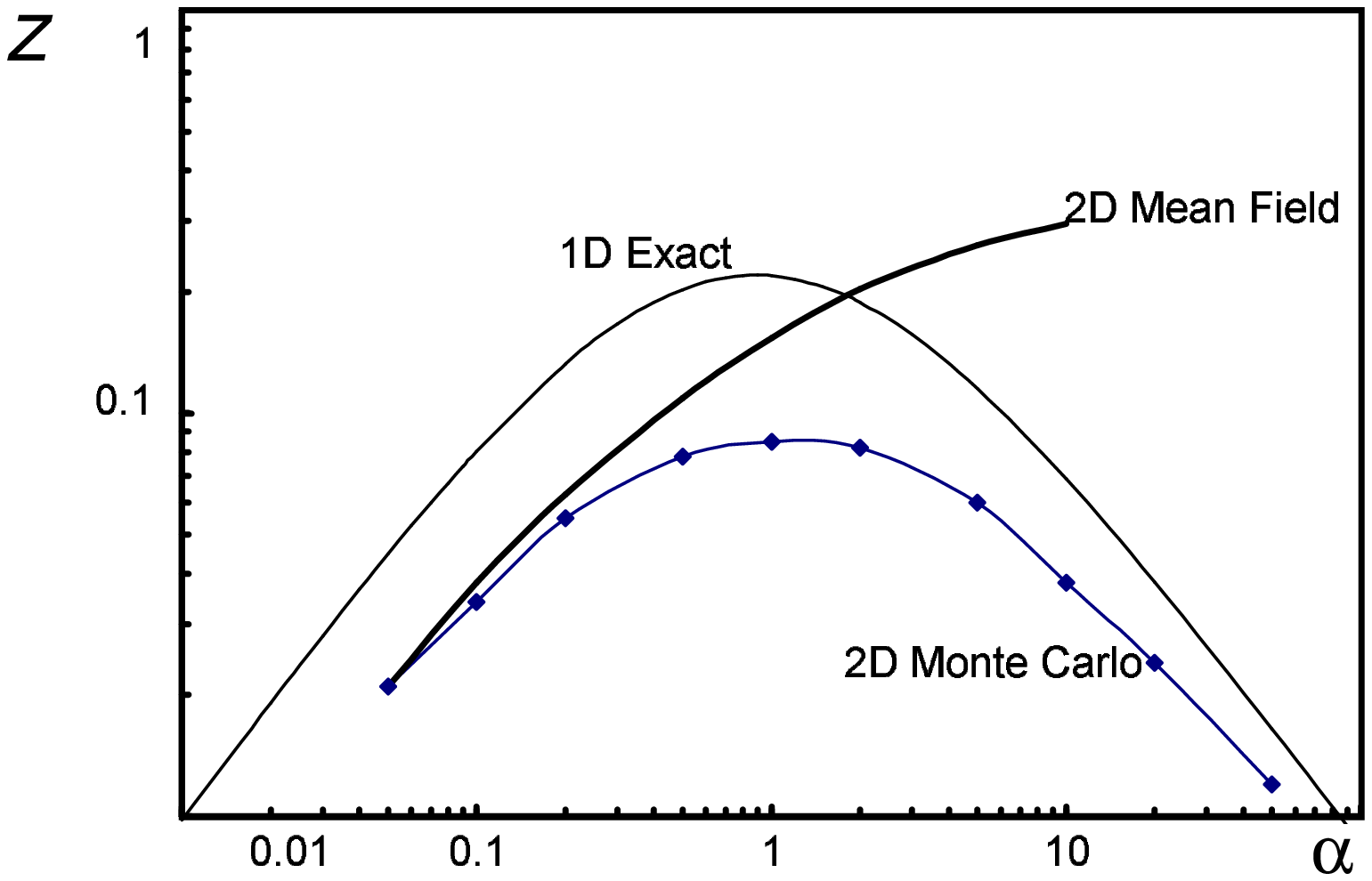}
\vskip 0.15in
\noindent{{\bf Fig.4} The kinetic exponent $z$ vs. $\alpha$
in one and two dimensions.}
\label{fig:4}
\end{figure}

\subsection{The Limit $\alpha\to\infty$}

When $\alpha=\infty$, the PRSA process develops through two stages.  The
initial stage is just RSA of monodisperse particles of radii $R=R_{\rm
max}=1$.  It continues until the jamming coverage, $\Phi_{\infty}\cong
0.542$, is reached.  Then the late stage begins, where the next disc to
place would be the one that fits the biggest hole.  The dynamics in this
late stage is thus deterministic and extremal.  This deterministic
procedure has been applied by Apollonius of Perga 200BC to fill the
space between three ``kissing'' discs\cite{mand}.  In the present case,
the procedure fills uncovered space obtained during the initial RSA
stage.  As the process develops, new discs will be placed more and more
often into the curvilinear triangles confined by three kissing discs.
Such curvilinear triangles are filled independently, so locally our
packing should be identical to Apollonian packing.  The fractal
dimension quantifies local characteristics of the pattern, so we
conclude that $D_f(\infty)=D_A$.  

The fractal dimension of Apollonian packing is $D_A\cong 1.3057$ in two
dimensions.  Note that the fractal dimension of the Apollonian parking,
arguably the oldest known fractal, has not been computed analytically in
higher dimensions, so the exact value of $D_f(\infty)=D_A$ remains
unknown.  The only trivial exception is the one-dimensional case where
the holes remaining after the initial stage are filled up completely
during the deterministic stage.  The number density $n(x)\equiv
C(x,t=\infty)$ of adsorbed intervals of length $x$ is\cite{mudak2}
\begin{equation}
\label{nx}
n(x)=2\int_0^\infty dt\,t\exp
\left[-xt-2\int_0^t d\tau{1-e^{-\tau}\over \tau}\right].
\end{equation}
$n(x)$ exhibits a weak integrable singularity in the small size limit,
$n(x)\sim \ln(1/x)$, implying that $N(\epsilon)$ is regular and therefore
$D_f(\infty)=0$\cite{mudak}.  This agrees with previous exact results,
Eqs.(\ref{eqbeta}) and (\ref{zbeta}), which provide the asymptotic behavior
$D_f(\alpha)\simeq {\ln 2\over \ln(\alpha+2)}$ when $\alpha\to\infty$.

\section{Ordering in the RSA}

For a quantitative description of the ordering processes we introduce an
entropy, $S_N$, characterizing the degree of order of $N$-particle
patterns.  Let $C$ denotes a particular $N$-particle pattern and
$p_N(C)$ denotes the probability of that pattern.  We can define the
Shannon entropy \cite{mackey,bla}:
\begin{equation}
S_N = -\sum_{C} p_N(C) \log_2 p_N(C)
\label{entdef}
\end{equation}
As it follows from Eq.~(\ref{entdef}), $S_N=0$ for a regular pattern,
since only one configuration (which occurs with the probability $p_N=1$)
contributes to the entropy.  A closely related quantity, $dS_N/dN \simeq
S_{N+1}-S_N$, gives the entropy production rate per particle and
characterizes the regularity of the pattern formation process. We also
introduce the conditional entropy, $ S_{N+1}^{*}(C)$, characterizing
patterns built by adding a disc to a given $N$-particle pattern $C$:
\begin{equation}
\label{entcond}
 S_{N+1}^*(C)=-\sum_{R, \vec{r}} 
p(R,\vec{r}\, |C)\log_2 p(R,\vec{r}\, |C).
\end{equation}
Here $p\left(R,\vec{r}\, |C\right)$ is the conditional
probability to add a disk of radius $R$ at point $\vec{r}$ to the
particular pattern $C(N)$ of $N$ disks. The total probability of
the $N+1$-particle configuration, obtained from the pattern
$C$ by placing an additional disc, reads:
\begin{equation}
\label{condprob}
p_{N+1}(R,\vec{r}\,,C)=p(R,\vec{r}\, |C) p_N( C).
\end{equation}
$S_{N+1}$ can be written as
\begin{equation}
\label{sn1}
S_{N+1}=
-\sum_{R, \vec{r},C}p_{N+1}(R,\vec{r}\,,C)
\log_2 p_{N+1}(R,\vec{r}\, , C).
\end{equation}
Using Eqs.~(\ref{entdef})--(\ref{sn1}) and the normalization condition,
$\sum_{R, \vec{r}}p(R,\vec{r}\,|C)=1$, we finally arrive at the entropy
production rate:
\begin{equation}
{dS_N\over dN} \simeq 
S_{N+1}-S_N=\sum_{C}p_N(C) S_{N+1}^*(C).
\label{entprav}
\end{equation}
Thus the entropy production rate is obtained by averaging the conditional
entropy over all possible configurations.

\begin{figure}
%\narrowtext
\epsfxsize=2.8in\epsfysize=2.4in
\hskip 0.3in\epsfbox{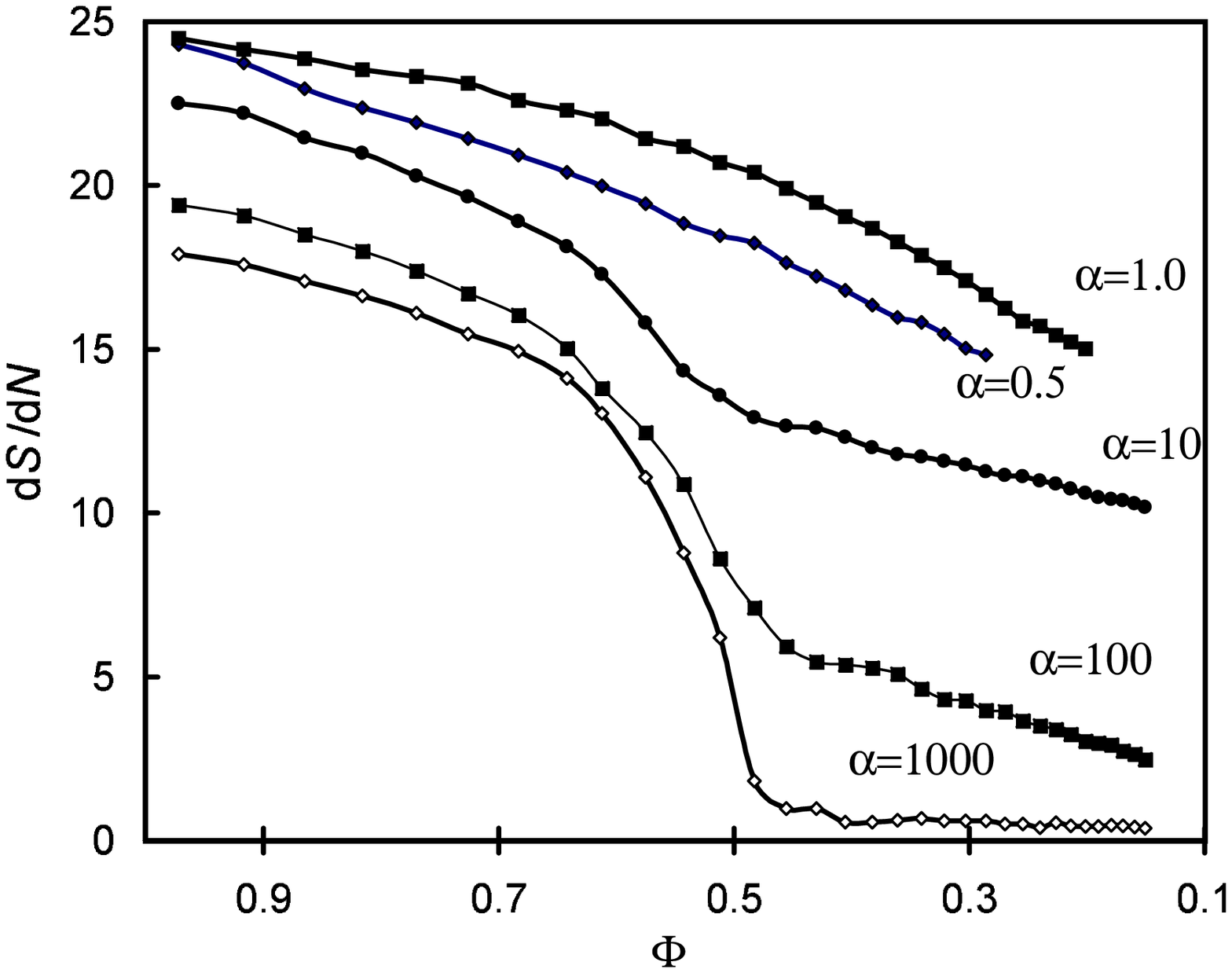}
\vskip 0.15in
\noindent{{\bf Fig.5} The entropy production rate vs. the
uncovered area for different values of $\alpha$.  A sharp decay
at $\Phi \approx 0.55$ is clearly seen for large $\alpha$.}
\label{fig:5}
\end{figure}

In spite of apparent simplicity of the definition of the entropy production
rate, its analytical evaluation is a very challenging problem even in one
dimension.  Indeed, it requires knowledge of the multiparticle probability
distribution function. In the simplest case of the monodisperse RSA we do
know the first two moments of $p(C)$, the density\cite{rev} and the pair
correlation function\cite{bon}, but much more detail information is needed
for determination of the entropy production rate.  Thus we investigated this
quantity numerically for 2D system by means of Monte Carlo simulations. We
implemented the following algorithm.  A disc which is added to the pattern
$C(N)$ of $N$ disks was treated as a point in a the ``configurational'' space
$(R, \vec{r})=(R,x,y)$.  We divided the continuous configuration space into
sufficiently small discrete cells and enumerated these cells. For
computations we used cells of linear size $0.01$ and $5 \times 5$ fragment of
the surface (with $R_{max}=1$, see (\ref{distrib})); thus, the total number
of cells was $M=2.5 \times 10^7$.  Then the conditional probabilities,
$p\left(R,\vec{r}\, |C(N)\right)=p_i$, that $(N+1)$-th disk comes to the
$i^{\rm th}$ cell, corresponding to $(R,\vec{r})$ in the given configuration,
$C(N)$, of $N$ discs was calculated numerically from
\begin{equation}
p_i=\frac{A_iR_i^{\alpha-1}}{\sum_{j=1}^M A_j R_j^{\alpha-1}}
\label{numcondp}
\end{equation}
where $A_i=0$ if the disc corresponding to the $i^{\rm th}$ cell overlaps
with some disc in the pattern $C(N)$; otherwise, $A_i=1$.
Eq.~(\ref{numcondp}) follows from the rate of adsorption, (\ref{distrib}),
and the normalization condition for the probability.  The discrete
conditional probabilities (\ref{numcondp}) are used to calculate the
conditional entropy $ S_{N+1}^*(C)$ via Eq.~(\ref{entcond}).  Finally, the
entropy production rate was obtained by averaging $S_{N+1}^*(C)$ over $C$
according to Eq.~(\ref{entprav}).  The averaging was performed over $10^2$
Monte Carlo runs, and the accuracy of the method was controlled through
run-to-run deviations.

To compare the entropy production rate for different values of $\alpha$,
we plot $dS_N/dN$ versus $\Phi$ (see Fig.~5).  The striking behavior of
the entropy production rate is clearly seen in the large $\alpha$ limit:
At the beginning of the process of pattern formation (i.e., when $\Phi
\approx 1$), the entropy production rate decreases slowly similar to the
small $\alpha$ case.  Around $\Phi \approx 0.55$, however, a sharp decay
occurs. The threshold value, $\Phi \approx 0.55$, is close to the
jamming density, $\Phi_{\infty}\cong 0.542$, of the RSA of identical
discs.  The transition demarcates the initial random stage when discs of
maximal radii are deposited and the final deterministic stage when discs
of maximal possible radius are inserted into maximal holes.  Given the
deterministic nature of the final stage, the entropy production rate
should be equal to zero when $\alpha=\infty$ and $\Phi \ge
\Phi_{\infty}$.  This is in agreement with our numerical findings for
large $\alpha$.

\section{Summary}

In summary, we investigated the adsorption kinetics and spatial properties of
the arising patterns in RSA of polydisperse particles whose size distribution
has a power-law form in the small size limit.  We developed a scaling
approach and verified that it indeed applies by comparing with exact results
in one dimension and numerical results in two dimensions.  We found that
arising patterns are self-similar fractals that appear to be completely
random when deposited particles are predominantly small; in the opposite
limit highly ordered structures, locally isomorphic to Apollonian parking,
are formed.  The fractal dimension $D_f$ of the pore space is determined by
the power-law exponent $\alpha$ of the particles size distribution.  When
$\alpha$ increases from $0$ to $\infty$, $D_f$ decreases from 2 to $D_A\cong
1.305$, of the Apollonian packing.  We introduced the entropy production rate
as a quantitative measure of the regularity of arising patterns.  We observed
that for sufficiently small $\alpha$, the entropy production rate smoothly
decays as the coverage increases.  In the complimentary case of $\alpha \gg
1$, the entropy production rate displays a similar behavior for sufficiently
small coverage, followed by a sharp decay to a very low entropy production
rate.  Physically, it reflects a two-stage nature of the pattern formation
process in the large $\alpha$ limit: The ordinary RSA, that goes until the
jammed state is reached, leads to a random structure which is starting point
for the second stage.  During this late stage the deposition process is
deterministic and extremal -- at a given step, the largest hole is filled.

\bigskip\noindent
One of us (PLK) acknowledges NSF grant DMR-9632059 and ARO grant
DAAH04-96-1-0114 for financial support.

\end{multicols}

\begin{thebibliography}{99}
\bibitem{rem102} 
     It was rigorously
     shown in \cite{pk} that for $\alpha \to 0$, the concentration of
     holes in the scaling limit, $y \to 0$, $t \to \infty$, behaves as
     $C(y,t) \sim t^2 \exp(-yt)$, i.e. the scaling function is
     exponential as in the mean-field theory.
\bibitem{rem5}
     $a_1=\frac{\pi^2}{6}-\frac{37\pi}{12} +\frac{\pi^2b_1}{3}-4\pi
     b_0$;\quad $a_2=\frac{\pi}{12}-\frac{\pi^2}{6}\left( 1+2b_1
     \right)$;\\ $a_3=\frac{19\pi^2}{12}-\frac{5a_2^2}{6\pi}
     -\frac{5a_2}{12}-2\pi b_0 -b_1\left(2a_2+2\pi a_2 b_3
     -\frac{7\pi}{2} \right)$;\\ $a_4=2\pi b_1+\pi
     +\frac{a_2}{2}-\frac12$;\quad $a_5=\pi -\frac12 +\frac{a_2}{2} + 2
     \pi b_1$. \\ The constants $b_k$ are given by $b_k=\int_0^{\infty}
     dx\,x^k \ln x\, e^{-\pi x^2}$ and may be expressed via gamma and
     digamma functions: $b_k={1\over 4}\,\Gamma({k+1\over
     2})\,\pi^{-(k+1)/2} \left[\psi({k+1\over 2})-\ln\pi\right]$.
\bibitem{bla}
     The summation over a continuous set of configurations implicitly
     assumes that we consider a finite subset of configurations;   
     in a numerical implementation, this is clearly the case.
\end{thebibliography}
\end{document}